\documentclass[manuscript]{acmart}

\usepackage{caption}
\usepackage{subcaption}

\AtBeginDocument{%
  \providecommand\BibTeX{{%
    \normalfont B\kern-0.5em{\scshape i\kern-0.25em b}\kern-0.8em\TeX}}}

\setcopyright{none}

\acmConference[FAccTRec '22]{5th FAccTRec Workshop: Responsible Recommendation}{September 23, 2022}{Seattle, WA}

\begin{document}

\title[The Users Aren’t Alright]{The Users Aren’t Alright: Dangerous Mental Illness Behaviors and Recommendations}

\author{Ashlee Milton}
\email{milto064@umn.edu}
\orcid{}
\author{Stevie Chancellor}
\email{steviec@umn.edu}
\affiliation{%
  \institution{University of Minnesota -- GroupLens Lab}
  \city{Minneapolis}
  \state{Minnesota}
  \country{USA}
}

\renewcommand{\shortauthors}{Milton and Chancellor}

\maketitle

\section{Introduction}

Recommendation systems permeate the technological ecosystem, from community engagement on social media to material needs from e-commerce sites. Often, the relationship between users and systems is presented as symbiotic, users are presented with tailored content while the system obtains data to improve recommendations. As researchers, we often take this cycle for granted and fail to consider users' well-being when designing recommendation systems.

So what happens when these recommendation systems present harmful or dangerous ideas to users, especially users with mental illness? Mental illness can manifest as serious self-harm, making the distribution and presentation of harmful ideas to such users problematic. Mainstream media has highlighted several instances in which assumptions of a symbiotic relationship fall apart. For example, Amazon has presented dangerous suicide cocktails \cite{twohey_dance_2022} and other systems have spread the Blue Whale challenge \cite{adeane_2019}. As many platforms use recommendations to disseminate content, recommendation systems are in essence facilitating the spread or ``contagion'' of this information through their platforms \cite{carlyle2018suicide, goldenberg2020digital, khasawneh2020examining}. Further, recommended content can lead to what we call ``algorithmic cruelty'', where algorithms directly harm or cause negative emotions in people. This has been well-studied with traumatic events like breakups \cite{pinter2019never} or death \cite{lustig2022designing}.

\textbf{In this paper, we argue that recommendation systems are in a unique position to propagate dangerous and cruel behaviors to people with mental illnesses.}  We hypothesize that recommendations risk exacerbating symptoms and behaviors which may unintentionally trigger individuals who are recovering or in relapse. We base this on prior work on how recommended content hurts people \cite{pinter2019never, lustig2022designing, carlyle2018suicide, goldenberg2020digital, khasawneh2020examining}, as well as three real examples of how recommendations hurt people with mental illness. We argue that recommendation system designers and researchers should consider the value of \textit{safety} of at-risk users, like those with mental illness. Finally, we propose future sociotechnical questions and discuss opportunities to improve this space for more fair, just, and equitable recommendations.

\section{Three Examples of Recommendation System Harms and Mental Illness}
In theory, recommendation systems internalize users with mental illness harmful behaviors and propagate them to other users. Recommendation systems are not human and lack the context of a given scenario - it has no concept of what is ``safe" or ``harmful". We begin by showcasing three examples of recommendation systems harms and explaining our concerns around their existence. 

\begin{figure}
\centering
\begin{subfigure}{.5\textwidth}
  \centering
  \includegraphics[width=\linewidth]{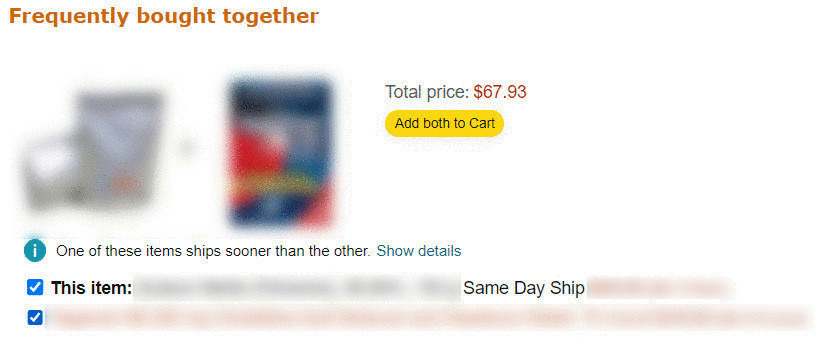}
  \caption{Chemicals A and B}
  \label{fig:amazon_suicide}
\end{subfigure}%
\begin{subfigure}{.5\textwidth}
  \centering
  \includegraphics[width=\linewidth]{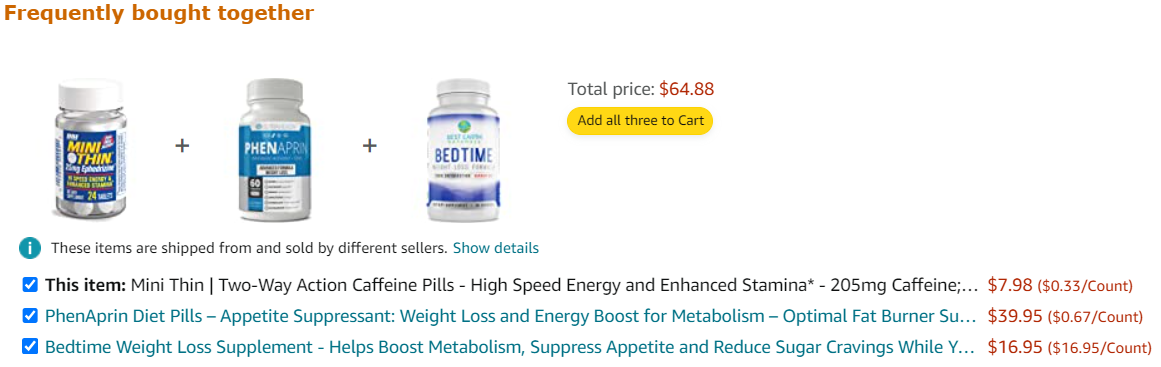}
  \caption{Caffeine Pills}
  \label{fig:amazon_caffine}
\end{subfigure}
\caption{Amazon Recommendations}
\label{fig:test}
\end{figure}

\textbf{Recommending Methods and Suicide Prevention:} We present Figure \ref{fig:amazon_suicide}, which represents the Amazon recommendation for Chemical A, a commercial substance commonly used by individuals to end their lives \cite{twohey_dance_2022} \footnote{We have anonymized this example to align with best practices for discussing suicide methods and suicide prevention.} The item recommended with Chemical A is Chemical B, another over-the-counter substance. When used together, these two substances counteract a person's reflexive self-preservation response and therefore increase the likelihood of death. While both Chemical A and B have practical uses individually, the combination is exclusively used as a method of suicide. Thus, the showcased recommendation is a one-click add-to-cart method for ending one's life. 

\textbf{Extreme Dieting.} While the previous example is explicit in its danger, we also found more subtle examples. Figure \ref{fig:amazon_caffine} depicts the recommendations associated with a caffeine supplement. The accompanying recommendations are for an appetite suppressant and a weight loss supplement. The combination of these substances, in theory, is ``logical''; however, dietary supplements are not reviewed by the FDA before they go to market \cite{FDA}, and there is no information provided on how the recommended supplements may interact. Moreover, when a person with an eating disorder encounters these recommendations, it can encourage harmful behaviors and abuse of substances to further extreme weight loss goals. 

\textbf{Blue Whale Challenge.} While the prior examples focus on e-commerce, problematic recommendations can be found on other platforms. A study on YouTube and Twitter found that the blue whale challenge, which encourages users to self-harm and kill themselves, showed signs of the contagion effect through posts and comments \cite{khasawneh2020examining}. Both platforms rely heavily on recommendations for distributing content to users. While harmful behaviors, and the products or content glorifying them, have always been present this is an example of how recommendation systems can expose at-risk users to harmful content.

\section{Big Questions and Future Work}
Based on these examples, we argue that recommendation systems are in a unique position to propagate dangerous and cruel behaviors to people with mental illnesses. These examples also lead to questions about personal safety as a value for recommendation systems. We present two questions for the community to consider and the potential impacts of future work on improving users' safety.

\subsection{Could recommender systems consider users' safety?}

One benefit of recommendation systems is their flexibility in adapting to users' needs. In the case of mental illness, this is a double-edged sword. Flexibility allows systems to quickly respond to changing needs, both from a user and corporate perspective, but creates opportunities for the introduction and propagation of unintentional harms, as we documented above. One starting place for addressing user safety is considering where harmful behaviors are being introduced and how can their effect on the system be mitigated.

Data is one starting place to look at aspects of safety. Recommendation systems are only as good as the data that they are provided and, of course, heavily rely on users' past behaviors and similar user behaviors. Data interventions would need to detect harmful behaviors so as to not let the system incorporate and then propagate them. There is existing work surrounding the detection of mental illness issues, including depression and suicide, in the broader field of information retrieval and social computing \cite{zhang2020detecting, zaman2019detecting, guntuku2017detecting}. Could we use these existing strategies to detect harmful user interactions? However, the use of detection systems is ethically complicated, as it intersects with privacy, data retention, and other ethical concerns. While these strategies could be a starting point, there should be serious ethical consideration of their use and future work should consider other strategies to improve safety. 

User behaviors alone are not the only place that could subvert user safety -- as we showed in our examples, pairs of recommendations make safety more than the sum of the proverbial parts. Recall the example from Figure \ref{fig:amazon_caffine} -- caffeine pills by themselves may not be unsafe; however, the combination of the supplements in the recommendation considers further thought. Is it safe for a recommendation system to suggest other products for medications and supplements when the system itself has no knowledge of possible side effects? Should weight loss supplements even be eligible to be a recommended product given their potential for harm? Examining the product or content that recommendation systems are allowed to recommend, we can subvert some of the potential harms to at-risk users.

Finally, context-aware and human-centered strategies for the evaluation of systems will also be crucial. Benchmarking metrics for recommendation systems prioritize performance, such as RMSE, MAE, or click-through rate. However, these metrics do not show how recommendations affect at-risk users. Simulation has been used to work around these problems \cite{kadian2020sim2real}, but is done in a general context that does not focus on vulnerable groups. Future work in context-aware and human-centered evaluation methods would open the door for better testing for at-risk users. 

\subsection{Who should be held accountable for user safety and how?}
Aside from the technical challenges we mention above, these recommendations prompt important questions about accountability and social responsibility. Who or what entity is responsible for the impacts of recommendations, if at all? To what extent can responsibility be enacted through social practices or law? It is not clear to what extent ``accountability'' or ``responsibility'' translates into an obligation to address harmful to at-risk users. Nor is it clear WHO is accountable for recommendations, many of which are not intentional and that no one would intend to make. 

We see some promising moves in the academic space to address this, taken from the focus on broader implications. For example, some academic conferences (like NeurIPS) now require a broader impact and ethical consideration section for submitted papers. These sections are a step in the right direction if more academic conferences adopt them. 

However, this means our best mechanisms for accountability only apply to publishing or post-hoc media attention (like the media attention on the Amazon and suicide methods example). How should accountability be exercised in this circumstance? The recommendation from that example can lead to the death of a person, which in some situations has legal consequences \cite{c.a.goldberg}. Even if there are no legal issues, is there civil or social recourse? There are deep ethical questions surrounding at-risk users' safety in recommendation systems that will need to be addressed as systems continue to propagate this behavior.


\section{Conclusion}
We have showcased how current recommendation systems are perpetuating harmful behaviors and posed questions with starting points that drive towards solutions. Dangerous recommendations related to mental illness are hard to detect and even more challenging to mitigate. Given these examples of recommendation systems and connections to mental illness, we must start considering users' safety or we risk physical harm to people. While user safety is the driving motivation, safety goes hand in hand with accountability and responsibility to respond to these concerns.

\bibliographystyle{ACM-Reference-Format}
\bibliography{sample-base}

\end{document}